\documentclass[aps,prd,10pt,letterpaper,preprintnumbers,floatfix,superscriptaddress,onecolumn]{revtex4}
\pdfoutput=1 
\usepackage{graphicx,dcolumn,bm,epsfig,cancel,hyperref}
\usepackage{multirow}
\usepackage{amsmath}
\usepackage{amssymb, times}
\usepackage[utf8]{inputenc}
\usepackage[normalem]{ulem}
\usepackage{color}
\usepackage{float}

\newcommand{\comment}[1]{{}}

\newcommand{\fastai}{\texttt{Fast.AI }}

\newcommand{\bea}{\begin{eqnarray}}  
\newcommand{\eea}{\end{eqnarray}}

\newcommand{\bsvar}{\sigma_{b}^2}
\usepackage{multirow}
\usepackage{booktabs}
\usepackage{makecell}
\usepackage{tabulary}
\usepackage{verbatim}
\usepackage{cancel}

\begin{document}
\title{
Towards recognizing the light facet of the Higgs Boson
}
\author{Alexandre Alves}
\email{aalves@unifesp.br}
\affiliation{Departamento de Física, Universidade Federal de São Paulo, UNIFESP, Diadema, Brazil}

\author{Felipe F. Freitas}
\email{felipefreitas@ua.pt}
\affiliation{CAS Key Laboratory of Theoretical Physics, Institute of Theoretical Physics,
Chinese Academy of Sciences, Beijing 100190, China}
\affiliation{Departamento de Física, Universidade de Aveiro and CIDMA, Campus de Santiago, 3810-183 Aveiro, Portugal}

\begin{abstract}
  The Higgs boson couplings to bottom and top quarks have been measured and agree well with the Standard Model predictions. Decays to lighter quarks and gluons, however, remain elusive. Observing these decays is essential to complete the picture of the Higgs boson interactions. In this work, we present the perspectives for the 14 TeV LHC to observe the Higgs boson decay to gluon jets assembling convolutional neural networks, trained to recognize abstract jet images constructed embodying particle flow information, and boosted decision trees with kinetic information from Higgs-strahlung $ZH\to \ell^+\ell^- + gg$ events. We show that this approach might be able to observe Higgs to gluon decays with a significance of around $2.4\sigma$ improving significantly previous prospects based on cut-and-count analysis. An upper bound of $BR(H\to gg)\leq 1.74\times BR^{SM}(H\to gg)$ at 95\% confidence level after 3000 fb$^{-1}$ of data is obtained using these machine learning techniques. 
\end{abstract}


\maketitle

\section{ Introduction}
The Standard Model (SM) Higgs boson established the spontaneous electroweak symmetry breaking (EWSB) mechanism as the responsible to give the particles their masses at the same time that it preserves the gauge symmetry of the SM interactions~\cite{PhysRevLett.13.508,PhysRevLett.13.321,PhysRevLett.13.585,PhysRev.145.1156,PhysRev.155.1554}. The immediate consequence of the EWSB mechanism is that the Higgs boson interactions scale with the particles masses $m$ as $\frac{m}{v}$, where $v\sim 246$ GeV is the vacuum expectation value of the Higgs field. Thus, interactions with heavier particles are likely to be observed first and that expectation has been fulfilled in the LHC where Higgs bosons $H$ interacting to $W$, $Z$, $b$, $\tau$ and $t$ have already been observed~\cite{Aad:2019mbh}. 

The couplings to heavy particles also permit the observation of 1-loop induced interactions of the Higgs boson to photons, gluons and $Z\gamma$ through couplings to top quarks and weak gauge bosons in the loop. In the case of photons, despite its tiny branching ratio, a narrow peak in the photons invariant masses on the top of a smooth monotonically decreasing background spectrum makes the observation possible in the gluon fusion production mode~\cite{Aad:2019mbh}. 

On the other hand, the coupling to gluons can just be inferred from the initial state gluon fusion into Higgs bosons~\cite{Englert:2014uua,Azatov:2013xha,Grojean:2013nya}. Contrary to photons pairs, the gluons pair decay is completely buried beneath a huge background of jet pairs from leading order QCD interactions turning its observation practically impossible in the gluon fusion channel. This motivates the search for untagged jets in Higgs decays, $H\to gg$ and $H\to q\bar{q}$, $q=u,d,s$, and also $b$ and $c$-jets, in cleaner production channels, for example, in the Higgs-strahlung process $pp\to VH,\; V=W,Z$. Such analyses have been carried out in Refs.~\cite{Carpenter:2016mwd,Perez:2015lra,Perez:2015aoa} and the prospects for the observation of the Higgs decay to light jets were found to be rather difficult, just an $1\sigma$ sensitivity after an integrated luminosity of 3000 fb$^{-1}$ and an upper bound of $BR(H\to jj)<4\times BR^{SM}(H\to gg)$, at 95\% CL, according to Ref.~\cite{Carpenter:2016mwd}. 
The authors of that work, however, suggest that a multivariate analysis might improve the sensitivity of the LHC searches compared to their standard cut-and-count analysis.

The Higgs decay to light jets is dominated by gluon decays. Even though $H\to gg$ is 1-loop induced, the Yukawa couplings of the Higgs with $u,d$ and $s$ quarks are suppressed and contribute too little to $H\to jj$ compared to the gluonic component. It is very important to confirm both the Higgs coupling to light quarks and to gluons. Despite the gluon coupling to Higgs can be inferred from inclusive gluon fusion processes, observing gluon pairs at the final states is necessary. First of all, it is known that even inclusive production rates very close to the SM can still hide new physics contributions~\cite{Schlaffer:2014osa}. These potential new physics contributions can be disentangled by measuring the final state kinematic distributions of the final state particles, thus it might be essential to measure the coupling directly at the decay of the Higgs boson. This way, observing $H\to gg$ directly makes it is easier to pin down the coupling strength and compare it with the SM expectation in the search for new particles running in the $Hgg$ loop.

There are many new physics scenarios that might contribute to the Higgs-gluon and Higgs-photon couplings at 1-loop, mainly those models that predict strongly interacting particles with SM electroweak charge as vector-like quark models~\cite{Aguilar-Saavedra:2013qpa}, squarks from supersymmetric extensions of the SM~\cite{Martin:1997ns},  and technicolor~\cite{Lane:1993wz}. Models that affect the Higgs-gluon coupling but not the Higgs-photon coupling are less common but exist as, for example, scalar color octets~\cite{GoncalvesNetto:2012nt} which, being electrically neutral, do not contribute to the Higgs-photon 1-loop coupling, but can couple to the Higgs boson as, for example, $H^\dagger H Tr(S^*S)$, where $S=\lambda^aS^a$, with $\lambda^a$ the Gell-Mann matrices, is an scalar color octet and electroweak singlet, and $H$ is the $SU(2)_L$ Higgs doublet. On the other hand, new physics scenarios might impact $H\to\gamma\gamma$ exclusively but not $H\to gg$, like vector-like leptons~\cite{Joglekar:2012vc}, extended scalar sectors like the two-Higgs doublet models~\cite{Branco:2011iw} and new $W$ bosons from left-right models~\cite{Mohapatra:1986}. In this later case, confirming that $H\to gg$ coincides with the SM expectation but $H\to\gamma\gamma$ provides valuable clues about the new interactions.

Another important motivation of being able to identify $H\to gg$ is having another Higgs channel available for searches. The branching ratio of the Higgs decay to gluons amounts to around 8\%, forty times larger than the branching ratio into photons. While $gg\to H\to gg$ remains hopeless at the LHC due the overwhelming QCD background, processes like Higgs-strahlung $pp\to VH$, $V=Z,W$, weak boson fusion $pp\to Hjj$ and double Higgs production $gg\to HH$ might benefit much with the new channel as source a of additional events. Of course, extracting information from this Higgs to light jets channel cannot be accomplished without an efficient way to identify them as yields of a Higgs boson decay. For this task, using multivariate tools is mandatory.

Following the suggestion of Ref.~\cite{Carpenter:2016mwd}, we use machine learning (ML) techniques in order to improve the prospects to observe gluon jet pairs from Higgs boson decays in the Higgs-strahlung process $pp\to Z(\to\ell^+\ell^-)H(\to gg),\; \ell=e,\mu$ at the 14 TeV LHC. Concerning the channel $ZH\to \ell^+\ell^- + gg$, the prospects of observation from Ref.~\cite{Carpenter:2016mwd} are even dimmer compared to the untagged jets case based on the combination of $ZH$ and $WH$ channels, reaching $0.25\sigma$ in the absence of systematic uncertainties.
Our results can be improved by combining the one and two-lepton categories from $ZH$ and $WH$ channels just like in the previous works of Refs.~\cite{Carpenter:2016mwd,Perez:2015lra,Perez:2015aoa} but, in the present study, it complicates the application of the ML techniques mainly due the increasing of signal and background categories for classification. In principle, this is not a major difficulty once the algorithms that we are going to use can handle this situation, however, the number of backgrounds increase considerably. Anyways, we will show that using just the two-lepton category is enough to considerably improve the prospects of the LHC to observe Higgs to gluon jets compared to previous, non-ML based analysis.

The discerning power of Convolutional Neural Networks (CNN) has been demonstrated in several studies where the separation of jet types is essential, especially in the gluon/quark-initiated jet separation and heavy quark tagging~\cite{deOliveira:2015xxd,deOliveira:2017pjk,Cogan:2014oua,ATL-PHYS-PUB-2017-017,Kagan:2016wnu,Komiske:2016rsd,Barnard:2016qma,Butter:2017cot,Pearkes:2017hku,Macaluso:2018tck,Diefenbacher:2019ezd}. Many of these investigations were performed using the classic jet image, that is it, a mapping of the energy deposits of jet constituents in cells of the $\phi\times\eta$ plane covered by the hadronic calorimeter (HCAL). In Ref.~\cite{Nguyen:2018ugw}, however, an ingenious manner to embody the particle flow information in the images was proposed -- representing leptons, photons and hadrons as geometrical figures whose sizes reflect their transverse energy deposits in the various segments of the detector. Including these extra pieces of information improves significantly the classification accuracy of the CNNs. Such an improvement showed itself essential in the present task of discerning jets from high $p_T$ Higgs bosons from the SM backgrounds in  $\ell^+\ell^- +jj$ events compared to the standard jet images approach. 

The backgrounds to $Z(\to \ell^+\ell^-)H(\to gg)$ involve $t\bar{t}$ and $Z$+jets events, among other subdominant sources, which far exceed the number of signal events at the LHC. 
Even after classifying QCD and non-QCD events with great accuracy with the help of the jet images, a large number of backgrounds still pollute the signals precluding an statistical significant observation of light jet decays. Instead of relying just on CNNs and jet image information, kinematic information from particles of the Higgs decays into leptons and jets can be used to further separate signals from backgrounds. The CNN information can be taken together with other available information in an ensemble of ML algorithms. In Ref.~\cite{Alves:2016htj}, stacking algorithms were shown to be very useful to classify Higgs bosons events at the LHC. We employed the same idea in this work with very good results, improving upon the previous results of Ref.~\cite{Carpenter:2016mwd}. 

Our paper is organized as follows, in Section~\ref{sec:model} we describe the signal and backgrounds simulations and cuts employed to define a signal region; Section~\ref{sec:const} is dedicated to explain how we construct the jet images that feed the CNN model; Section~\ref{sec:CNN_model} then contains the details of the construction of the CNN model and its training methodology. The performance of the ML algorithms are discussed in Section~\ref{sec:RESULTS}, while Section~\ref{sec:aaet} presents our results. The conclusions can be found in Section~\ref{sec:conclusions}.

\section{\label{sec:model} Signal and Backgrounds in the $ZH\to \ell^+\ell^- + jj$ channel}
The most promising production mechanism for the identification of a hadronic decaying Higgs is the associated production with a gauge boson ($W$/$Z$) decaying to both charged leptons and neutrinos, while 
the Higgs boson can be detected using the reconstructed invariant mass of the hadronic products. Including $W$ bosons in the decay chain increases the number of signal events but it also makes the separation from huge backgrounds like $t\bar{t}$ and $W$+jets more challenging. We choose to allow just the leptonic $Z$ decays to facilitate the identification of our signals. The Higgs branching ratio to gluons is $\sim 360$ times larger than the three light quarks $u,d,s$ combined for a 125 GeV Higgs boson. We thus perform simulations of the following signal processes
\begin{equation}
    q\bar{q}, gg \rightarrow Z(\to \ell^+\ell^-)\; H(\to gg),\; \ell=e,\mu\; .
\end{equation}

The $gg \rightarrow ZH$ contributes to up to 20\% to the total rate of $ZH$ production. All the events for $q\bar{q} \to ZH$ were generated using \texttt{MadGraph}~\cite{Alwall:2014hca} at the leading order, with the Higgs effective coupling model and NN23NLO PDFs~\cite{Ball:2013hta}. For the $gg \rightarrow ZH$ process with quark loops, we used the \texttt{Madspin} module~\cite{Artoisenet:2012st} to generate the decays of the $Z$ boson into a pair of leptons and $H$ into our light jets ($H\rightarrow gg$) signal as well the $H\rightarrow b\bar{b}$ and $H\rightarrow c\bar{c}$ backgrounds.  
We apply overall rescaling QCD K-factors to the signal and background processes to match the total NNLO QCD and NLO EW cross section results taken from the Higgs cross section working group~\cite{deFlorian:2016spz}.

We generated 450000 signal events and around 2 million background events distributed amongst the background classes. All the events were showered and hadronized using \texttt{PYTHIA8}~\cite{Sjostrand:2007gs}. Hadronized events were then passed to \texttt{DELPHES3} \cite{deFavereau:2013fsa} to simulate detector effects. In order to build jet substructures for image classification and also to facilitate the vetoing of background jets, we search for the fat-jets reconstructed with radius of $R=1.5$ and $p^{min}_{T} =75$ GeV. In addition we turn on the computation of N-subjettiness variables using optimized (one-pass) anti-kt with $\beta = 1$. 
The backgrounds considered comprise the following irreducible and reducible ones: (1) $Zj(jj) \rightarrow \ell^{+}\ell^{-} + j(jj)$ up to two extra jets with the MLM matching scheme~\cite{Hoche:2006ph}, (2) $ZZ \rightarrow \ell^{+}\ell^{-} + jj$, (3) $WZ \rightarrow \ell^{+}\ell^{-} + jj$, (4) $t\bar{t} \rightarrow \ell^{+}\ell^{-} + \nu_{\ell}\bar{\nu}_{\ell} + jj$, (5) $Wj(jj)\to \ell^{+}\ell^{-} +\hbox{jets}$ where the second lepton comes from jet misreconstruction. These backgrounds are efficiently suppressed by the cuts of Eq.~\eqref{eq:cutI}--\eqref{eq:cutV} below and can be safely neglected.
We also included the decays of (6) $H \to b\bar{b}$ and (7) $H \to c\bar{c}$ in the Higgs-strahlung $q\bar{q}, gg \rightarrow ZH$ processes to take into account the mistagging event contamination. There are six background classes in total effectively.

\begin{table*}[t!]    
\centering
\begin{tabular}{l|c|c|c|c|c|c}
\toprule
\hline
Process cross section (fb) & \makecell{Basic Selection \\ Eq.~\eqref{eq:basic_cuts}} & Eq.~\eqref{eq:cutI}  & Eq.~\eqref{eq:cutII} & Eq.~\eqref{eq:cutIII} & Eq.~\eqref{eq:cutIV} & Eq.~\eqref{eq:cutV}\\
\hline
\hline
\midrule
\makecell{$q\bar{q}\rightarrow ZH,Z\rightarrow \ell^{+}\ell^{-}, H\rightarrow g g$ \\
$gg\rightarrow ZH,Z\rightarrow \ell^{+}\ell^{-}, H\rightarrow g g$}& \makecell{$3.87\times 10^{-1}$ \\ $1.17\times 10^{-1}$} & \makecell{$1.59\times 10^{-1}$ \\ $3.51\times 10^{-2}$} & \makecell{$2.99\times 10^{-2}$ \\ $1.54\times 10^{-2}$} & \makecell{$1.63\times 10^{-2}$ \\ $5.70\times 10^{-3}$} & \makecell{$1.57\times 10^{-2}$ \\ $5.20\times 10^{-3}$} & \makecell{$1.48\times 10^{-2}$ \\ $4.70\times 10^{-3}$} \\
\hline
\makecell{$q\bar{q}\rightarrow ZH,Z\rightarrow \ell^{+}\ell^{-}, H\rightarrow b\bar{b}$ \\
$gg\rightarrow ZH,Z\rightarrow \ell^{+}\ell^{-}, H\rightarrow b\bar{b}$}& \makecell{$1.25\times 10^{1}$ \\ 3.45}& \makecell{$5.31$ \\ $1.044$} & \makecell{$9.79\times 10^{-1}$ \\ $4.18\times 10^{-1}$} & \makecell{$4.84\times 10^{-1}$ \\ $1.42\times 10^{-1}$} & \makecell{$4.69\times 10^{-1}$ \\ $1.29\times 10^{-1}$} & \makecell{$4.28\times 10^{-1}$ \\ $1.09\times 10^{-1}$} \\
\hline
\makecell{$q\bar{q}\rightarrow ZH,Z\rightarrow \ell^{+}\ell^{-}, H\rightarrow c\bar{c}$ \\
$gg\rightarrow ZH,Z\rightarrow \ell^{+}\ell^{-}, H\rightarrow c\bar{c}$}& \makecell{1.24\\ $3.66\times 10^{-1}$}& \makecell{$5.28\times 10^{-1}$ \\ $1.10\times 10^{-1}$} & \makecell{$1.10\times 10^{-1}$ \\ $5.01\times 10^{-2}$} & \makecell{$6.14\times 10^{-2}$ \\ $1.83\times 10^{-2}$} & \makecell{$5.95\times 10^{-2}$ \\ $1.67\times 10^{-2}$} & \makecell{$5.49\times 10^{-2}$ \\ $1.46\times 10^{-2}$} \\
\hline
$Z+j(jj) \to \ell^{+}\ell^{-} + j(jj)$ &
$2.12\times 10^5$ & $8.83\times 10^4$ & $7.85\times 10^3$ & $1.02\times 10^3$ & $9\times 10^2$ & $8.02\times 10^2$ \\
\hline
$ZZ \rightarrow \ell^{+}\ell^{-} + jj$ & $1.31\times 10^2$ & $5.31\times 10^{1}$ & $5.07$ & $1.09$ & $1.02$ & $9.21\times 10^{-1}$ \\
\hline
$WZ \rightarrow \ell^{+}\ell^{-} + jj$ & $1.44\times 10^2$ & $6.41\times 10^{1}$ & $7.54$ & $1.05$ & $9.79\times 10^{-1}$ & $8.83\times 10^{-1}$ \\
\hline
$t\bar{t} \rightarrow \ell^{+}\ell^{-} + \nu_{\ell}\bar{\nu}_{\ell} + b\bar{b}$ & $7.52\times 10^3$ & $1.48\times 10^{3}$ & $2.12\times 10^{2}$ & $3.77\times 10^{1}$ & $1.54\times 10^{1}$ & $3.41$\\
\hline
\hline
\end{tabular}
\caption{Cross sections, in fb, for each signal and background process after successive selection criteria of Eqs.~\eqref{eq:basic_cuts}--\eqref{eq:cutV}.}
\label{tab:sele_cuts}
\end{table*}

In order to stay safely away from infrared and colinear divergences, we apply the basic cuts of Eq.~\eqref{eq:basic_cuts} at the generation  level
 \begin{eqnarray}
 && p^{\ell}_{T} > 20\; \hbox{GeV},\; p^{j,b}_{T} > 30\; \hbox{GeV},\nonumber\\  
 && \vert\eta^{j,b}\vert < 3.0,\; \vert\eta^{\ell}\vert < 2.7,\nonumber \\
 && \Delta R_{jj,bb,\ell\ell} > 0.01\; .
 \label{eq:basic_cuts}
 \end{eqnarray}

Beside the basic cuts we also imposed the following cuts, inspired in Ref.~\cite{Carpenter:2016mwd}, to further eliminate backgrounds and select boosted Higgs bosons
\begin{eqnarray}
&& \hbox{at least two same-flavour opposite-charge leptons with:}\nonumber\\
&& |\eta_{\ell}| < 2.5,\; p^{\ell}_{T} > 30\; \hbox{GeV},
\label{eq:cutI} \\
&& \hbox{at least one central jet with:}\nonumber\\
&& |\eta_j| < 2.0,\; p^{j}_{T} > 150\; \hbox{GeV},
\label{eq:cutII} \\
&& |M_j-m_H|<20\; \hbox{GeV}\label{eq:cutIII}, \\
&& M_{\ell\ell} > 80\; \hbox{GeV},\; p^{\ell\ell}_{T} = (p^{\ell_{1}}_{T} + p^{\ell_{2}}_{T}) > 100\; \hbox{GeV}\label{eq:cutIV},\\
&& \not\not\!\! {E_T} < 40\; \hbox{GeV},\label{eq:cutV}
\end{eqnarray}
where $M_j$ is the invariant mass of the leading fat-jet, $m_H=125$ GeV is the Higgs mass, $M_{\ell\ell}$ is the mass of the two leading charged leptons, $p^{\ell_{1}}_{T}$ and $p^{\ell_{2}}_{T}$ are the transverse momentum of the leading and the subleading leptons, respectively, and $\not\not\!\! {E_T}$ is the missing transverse energy of the event. 

In Table~\ref{tab:sele_cuts}, we show the cut flow of the cross sections for each signal and background channel for these selections and the basic selection of Eq.~\eqref{eq:basic_cuts}. The Higgs mass window of Eq.~\eqref{eq:cutIII} is very efficient in eliminating backgrounds without a Higgs boson while Eq.~\eqref{eq:cutV} is effective to veto $t\bar{t}$ events that contain neutrinos. Requiring a very hard leading jet, Eq.~\eqref{eq:cutII}, helps to decrease the backgrounds of weak bosons with extra jets. The cut on the mass of the leptons pair, $M_{\ell\ell}$, helps to eliminate backgrounds where the leptons are not produced from a $Z$ boson decay. All these kinematic variables are used to feed a BDT algorithm as we are going to discuss in section~\ref{sec:RESULTS}. The BDT is able to find even more efficient cut criteria that will help to separate signals and backgrounds events further.

We see that these cuts reduce the $Z$+jets and $t\bar{t}$ backgrounds of more than $\sim 2$ orders of magnitude, however, they are still 4 orders of magnitude larger than the signals. At this level of cuts only around 50 signal events are surviving after 3000 fb$^{-1}$. Hardening cuts will not make a better job, so we need a better plan to select the signal events. This strategy must focus mainly in the most dangerous backgrounds of $Z$+jets and $t\bar{t}$. We discuss next the selection tool that we are going to use to raise the prospects to identify gluon decays of the Higgs boson.

\section{\label{sec:const} Construction of abstract images -- embodying information to boost the discrimination power}
The use of computer vision and jet-images \cite{Cogan:2014oua,deOliveira:2015xxd} have proved their places among the most powerful and efficient techniques for classifying data according to different hypothesis \cite{Jung:2019iii,ATL-PHYS-PUB-2017-017,Kagan:2016wnu,Komiske:2016rsd,Barnard:2016qma,Butter:2017cot,Pearkes:2017hku,deOliveira:2017pjk,Macaluso:2018tck}. Although the use of jet-images is a very well established framework, in some cases the information contained in jet-images is not enough to deep convolutional neural networks to make reliable predictions. Cases where either the differences between signal and background are very subtle or the signal in question occur in a much smaller rate than the background events (i.e. rare events) might be difficult to disentangle. Different approaches to increase the amount of information contained in a jet-image have been proposed~\cite{Komiske:2016rsd,Diefenbacher:2019ezd,Kim:2019wns}. These methods employ a hybrid use of high-level features (i.e. kinematic observables) together with the information recovered from each detector section (electromagnetic calorimeter (ECAL), hadronic calorimeter (HCAL), Muon chambers, tracking system) encoded as image channels, greatly improving the amount of information, which subsequently increase the discriminant power of the algorithms. However, this improvement comes at the price of drastically increase of the model complexity which leads to overfitting  and/or a slow training phase for the NN. A simpler approach was proposed in Ref.~\cite{Nguyen:2018ugw} to tackle this problem using abstract shapes as a method to encode the information from the particles detected in each event. 

\begin{figure}[!t]
\begin{center}
\includegraphics[width=0.3\textwidth]{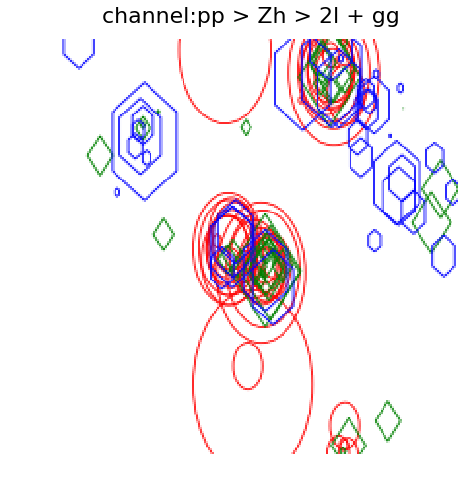} 
\includegraphics[width=0.3\textwidth]{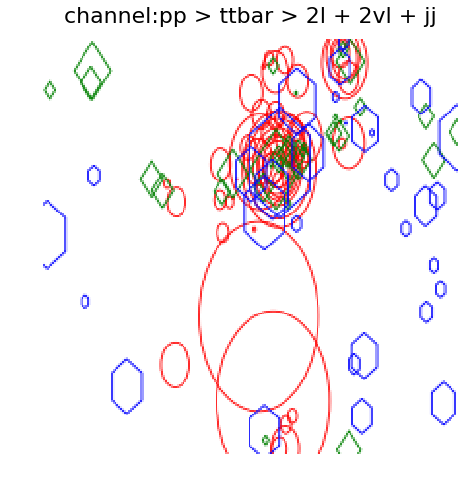} 
\caption{Left panel: abstract image for a signal event. Right panel: abstract image for a $t\bar{t}$ event. Charged particles (red circles), neutral hadrons (blue hexagons) and photons (green squares) are displayed in the $\eta$ (horizontal axis) {\it versus} $\phi$ (vertical axis) plane.} 
\label{fig_abstract}
\end{center}
\end{figure}

After the event generation, reconstruction and selection, we can construct the so called {\it abstract images} by following the guidelines of the Ref.~\cite{Nguyen:2018ugw}. \texttt{Delphes} uses a particle-flow algorithm~\cite{deFavereau:2013fsa} which produces two collections of 4-vectors \textemdash particle-flow tracks and particle-flow towers \textemdash that serve later as input for reconstructing high resolution jets and missing transverse energy. In order to construct the abstract images, we use the information stored in the particle-flow tracks. These tracks outputs from the \texttt{Delphes} particle-flow algorithm are stored in the \texttt{EFlowTrack} array which are used to construct the objects displayed as red circles in Fig.~\eqref{fig_abstract}. \texttt{EFlowPhoton} and \texttt{EFlowNeutralHadrons} store the photon and neutral hadrons output, respectively, which, by their turn, are used to construct the objects displayed as as green squares and blue hexagons, respectively, in Fig.~\eqref{fig_abstract}. All the shapes are centered at the $\eta\times\phi$ coordinates of the object (charged particles, photons or neutral hadrons) and their radius are proportional to the logarithm of their transverse momentum. In Figure~\eqref{fig_abstract} we can see an abstract image produced from an event for the signal $ZH\to \ell^+\ell^- + jj$ and the $t\bar{t}$ background.
 
%
\begin{table*}[ht!]
\centering
\begin{tabular}{c|c|c|c}
\toprule
\hline
    Process & Simulated & Synthetic & Total \\
    \hline\hline
    $ZH\to \ell^+\ell^- +gg$ &  35280 & 0 & 35280 \\
    \hline
    $ZH \to \ell^+\ell^- + b\bar{b}$ & 63341 & 422 &  63763\\
    \hline
    $ZH \to \ell^+\ell^- + c\bar{c}$ & 79977 & 0 & 79977 \\
    \hline
    $Z+j(jj)\to \ell^+\ell^- +j(jj)$ & 10158 & 25222 & 35380 \\
    \hline
    $ZZ\to \ell^+\ell^- +jj$ & 6673 & 26837 & 33510\\
    \hline
    $WZ\to \ell^+\ell^- +jj$ & 5822 & 27769 & 33591 \\
    \hline
    $t\bar{t}\to \ell^+\ell^- +\nu_\ell\bar{\nu}_\ell+b\bar{b}$ & 1129 & 29797 & 30926 \\
    \hline
    total & 202380 & 110047 & 312427\\
    \hline\hline
\end{tabular}
\caption{The number of simulated and synthetic images for each event class. The last column shows the total number of images adding the simulated and synthetic data. The number of synthetic data were generated in order to get a more balanced number of instances across the classes. The larger number of $\ell^+\ell^- b\bar{b}$ and $c\bar{c}$ instances do not represent an issue for the training of the algorithms.}
\label{tab:class_imagens}
\end{table*}


 The abstract image data set consists of 8-bit/color RGBA jet images of resolution $224\times 224$ pixels.
 We refer to this original data set as {\it simulated images}.
 After applying the selection criteria of Table~\ref{tab:sele_cuts}, we get a very imbalanced data set which is reflected in the relative number of simulated images of each class displayed in Table~\ref{tab:class_imagens}. The low number of samples for some of the background channels are due to the selection criteria that we imposed, which drastically reduces the number of selected events and consequently the number of images even after the simulation of $\sim 2.5$ millions of signal and background collision events at the LHC. It is important to point out that without demanding hard cuts it would be virtually impossible to identify the signals even with a very powerful ML classifier. From Table~\ref{tab:sele_cuts}, we see that the signal to background ratio is of order $10^{-6}$ before cuts.
 
 In order to balance the dataset across the classes and make sure that the CNN model will not overfit towards the classes with more instances, we can either use class weights to overcome this issue or generate more images "artificially" (data augmentation) for the classes where we have fewer samples. Actually, we augmented all the classes but the signal and the $ZH\to\ell^-\ell^+ +c\bar{c}$ classes which are those ones with the larger number of simulated images in our analysis. We refer to this artificially augmented data set as {\it synthetic images}. Although the class weighting is the less computationally expensive and straightforward approach, transforming the images to create new ones is a well tested approach commonly used in image recognition tasks where collecting large image datasets is infeasible. Moreover, it is important to emphasize that, as long as the test set contains only simulated images, the train set can be populated with anything that can help to control overfitting and improve the performance of the algorithm on the test set. 
 

We thus should emphasize that we generate the artificial images only for the training phase. For the evaluation/test phase we used simulated images strictly. 
The synthetic images are generated according to the following data augmentation scheme:
\begin{itemize}
    \item randomly select an image sample from the class we want to augment,
    \item resize and crop the image,
    \item rotate the image in a range of -30 to 30 degrees,
    \item horizontally flip,
    \item apply random noise. 
\end{itemize}

Each transformation has an uniform probability to be applied in a selected image, i.e. for a random selected original image, we generate a new artificial image by randomly selecting one of the 5 transformations listed above. In order to generate more artificial images without take the risk of over-saturate the data set with just copied images, we allow the selected images to have two or more transformations applied in the same image, with the caveat that each new transformation has a 50\% chance to be applied. For example, for a given selected image, we have 100\% chance of one of the 5 transformations be applied, a second transformation has 50\% chance of applied in the same image, a third transformation has 25\% chance, a fourth transformation will have 12.5\% chance of be applied, and so on. We also ensure that the transformations are not applied two times consecutively in the same image, so that if an image has first an horizontal flip, the next transformation will not be flipped horizontally again.


\section{\label{sec:CNN_model} CNN architecture and training methodology}
In this section, we describe in detail our training methodology. We tested several procedures and strategies to achieve robust and reliable results paying special attention in estimating the uncertainty of our results which will be provided in the next section. Some of these methods reflect the state-of-art in machine learning and our work serve as a good test in a particle physics application as well.

We want to classify whether a given abstract image belongs to one of the 7 classes: the signal class $ZH(jj)$, and the background classes $ZZ,WZ,Z+j(jj),ZH(b\bar{b}),ZH(c\bar{c}),t\bar{t}$. 
For this task we choose the Residual Network (ResNet) as the base architecture. ResNets were first proposed in Ref.~\cite{HeZRS15} and consist of a deep neural network (DNN) built as blocks of convolutional layers together with short cut connections (or skip layers) that help the ResNet to avoid problems associated with DNN, in particular, the well known gradient vanishing/exploding problem~\cite{279181}. In our analysis, we tested the discriminant power of the ResNet with an increasing number of layers: ResNet-18, ResNet-34, ResNet-50, and ResNet-101, using the abstract images data set described in the Section~\ref{sec:const}. The deeper configurations, ResNet-50 and ResNet-121, presented a final accuracy larger than 90\% while the shallower configurations, ResNet-18 and ResNet-34, reached the 65--70\% mark in the test set. The shallower configurations could be trained in less time than the deeper ones. The time to train and tune a ResNet-121 is, by its turn, much larger than ResNet-50 and, taking in consideration the training time, classification accuracy and signal significance, we chose ResNet-50. 

The ResNet-50 consists of 50 convolutional (Conv2D) layers, in between each Conv2D layer we have a series of batch normalizations, average pooling and rectified activations (ReLU). For our task, we replace the last fully connected layers of the ResNet-50, responsible for the classification, with the following sequence of layers:

\begin{itemize}
\item  an adaptive concatenate pooling layer ({\it AdaptiveConcatPool2d}), 
\item  a flatten layer,
\item  a block with batch normalization, dropout, linear, and ReLU layers,
\item  a dense linear layer with 7 units as outputs, each unit corresponding to a class and a softmax activation function.
\end{itemize}

The {\it AdaptiveConcatPool2d} layer uses adaptive average pooling and adaptive max pooling and concatenates them both. Such procedure provides the model with the information of both methods and improves the overall performance of the CNN. We also make use of the label smoothing methodology~\cite{DBLP:journals/corr/SzegedyVISW15} as well as the MixUp~\cite{DBLP:journals/corr/abs-1710-09412} training method, more details about these two techniques are presented in the Appendices~\ref{labelsmoothie} and ~\ref{mixup}.

One important aspect of the training of DNN models, and yet often not given the due attention, is the choice of the batch size. The use of large batch sizes helps the optimization algorithms to avoid overfitting \cite{DBLP:Samuel,DBLP:Smith15a, Akiba2017ExtremelyLM} acting as a regularizer. However, the batch size is ultimately bounded by the amount of memory available in hardware. One way to work around this limitation is the use of mixed precision training~\cite{DBLP:Micikevicius}, this method uses half-precision floating point numbers, without losing model accuracy or having  to modify hyper-parameters. This nearly halves memory requirements and, on recent GPUs, speeds up arithmetic calculations. 

The learning rate and weight decay are other two key hyperparameters to train DNNs. A good choice of these two parameters can greatly improve the model performance, in our particular case it means a high accuracy classification and good signal significance, and reduce drastically the training time. Instead of using a fixed value for the learning rate we opted to use the so called Cyclical Learning Rates (CLR)~\cite{DBLP:Smith15a}.  To use CLR, one must specify minimum and maximum learning rate boundaries and a step size. The step size is the number of iterations used for each step and a cycle consists of two such steps – one in which the learning rate increases and the other in which it decreases~\cite{DBLP:Smith15a}.

Following the guidelines from Ref.~\cite{DBLP:1803-09820}, we perform a scan over a selected range of values for learning rates and weight decays. 
According to Ref.~\cite{DBLP:1803-09820}, the best initial values for learning rates are the ones who give the steeper gradient towards the minimum loss value, which in our case, was found to be $\sim 3.0\times 10^{-5}$ for the learning rate and $1.0\times 10^{-5}$ for the weight decay, and for the maximum learning rate value we just multiply the initial value by 10. The particular architecture of the dense layers of the ResNet-50 and the hyperparameters were found using a genetic algorithm. This genetic algorithm is capable to find the best combination of hyperparameters and neural networks modules (a.k.a. layers) which maximize the significance of our signal.


From the 312k images displayed in Table~\ref{tab:class_imagens}, 187k (60\%) were used to train and test the CNN. All the synthetic images are used in this stage of the analysis. These 187k images were randomly split in 80\% to train and 20\% to test the CNN algorithm.

We trained our model in a 3-stage scheme: in the first stage, we trained the ResNet-50 all the way from the beginning to the end within 50 epochs without the use of transfer learning. At the end of the 50 epochs, we save the weights and bias of the trained model. Next, in the second stage, we loaded these weights from the previous trained model and "freeze" all layers up to the last 3 ones, so that during this training phase the backpropagation only takes effect on the parameters of the last 3 layers (this methodology is what we call {\it transfer learning}), we then trained these last 3 layers for 25 epochs. In the last stage, we loaded the weights and bias saved from the stage two and "freeze" all layers up to the last layer (the classification layer or header) and trained for 15 epochs. This 3-stage training scheme help us to find the most stable results while gradually increasing the performance of our model. In order to complete the training, an NVIDIA GTX 1070 GPU took 24 hours approximately. Running the tuning of hyperparameters with the genetic algorithm takes an amount of time three times larger. 

The CNN architecture, training methodology and all the state-of-art techniques employed, such as mixed precision training and MixUP, as well as the tuning of the hyperparameters were all done using Pytorch~\cite{NEURIPS2019_9015} and the \fastai~\cite{fastai} framework. The \fastai framework enabled us to easily implement all the techniques described above and made possible to modify all the aspects from the ResNet-50 with very few lines of codes.
\section{\label{sec:RESULTS} Performance of the classifiers}

\subsection{Classification with ResNet-50}
\begin{figure}[!t]
\begin{center}
\includegraphics[scale=0.35]{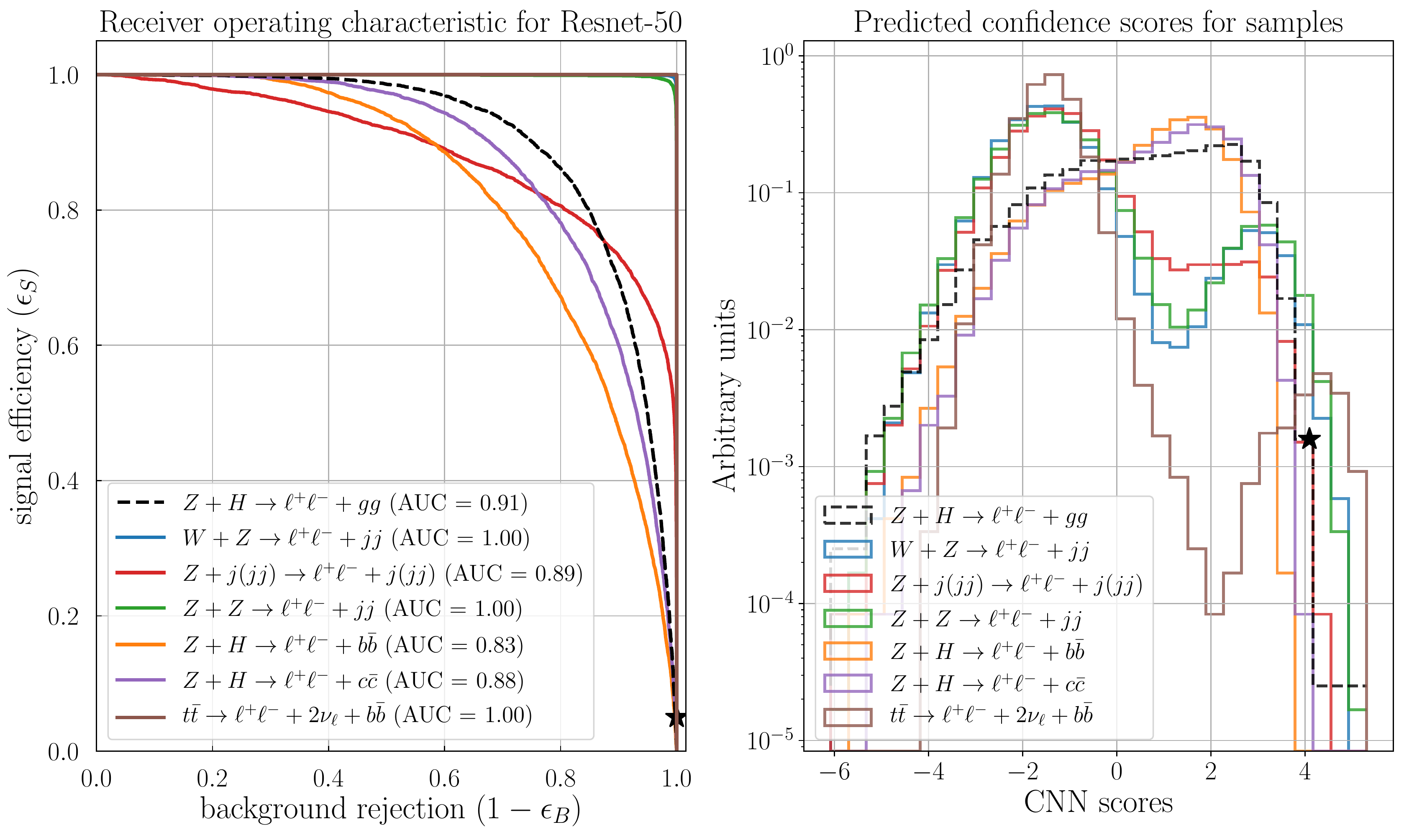} 
\caption{Left panel: ROC curves for the signal and backgrounds classes. Right panel: CNN scores for the signal and backgrounds classes. The star marker shows the cut selection that we impose to achieve the best Asimov significance given by this method.}
\label{ROC_1}
\end{center}
\end{figure}

The performance of our training framework of the CNN with abstract jet images can be evaluated on the basis of signal efficiency and background rejection factors as shown at the left panel of Fig.~\eqref{ROC_1} for each one of the seven event classes. The more rectangular is the Receiving-Operator-Characteristic (ROC) curve, the more efficient is the background rejection for a fixed signal acceptance. A simple figure to evaluate how good is the signal-background separation is the area under the ROC curve, AUC. The closer AUC is to one, the better we should expect the backgrounds can be cleaned up for a giving signal efficiency. In order to construct the ROC curves, we impose cuts on the CNN scores of the classes displayed at the right panel of Fig.~\eqref{ROC_1} and compute the number of events left for the signal versus a chosen background for which we want to obtain the ROC curve. The ideal separation is having the CNN score distribution of backgrounds concentrated at the left and the signal at the right in such a way that a cut on the scores eliminates more backgrounds than signal events. The range of the CNN scores seen at the right panel of Fig.~\eqref{ROC_1}, roughly from -6 to +6, is a feature of the particular loss function used to train the ResNet, see Appendix~\ref{labelsmoothie} for more details. 

From the ROC curve shown in Fig.~\eqref{ROC_1}, we can choose the point with the highest
signal significance which depends on the integrated luminosity and also on the effect of systematic uncertainties which are often disregarded in machine learning studies. The Asimov estimate of significance~\cite{Cowan:2010js}, a well-established approach to evaluate likelihood-based tests of new physics taking into account the systematic uncertainty on the background normalization, can then be used for a more careful estimate of the signal significance at the training and testing phases of construction of the classifier. The formula of the Asimov signal significance is given by
\begin{equation} \label{eq:asimov}
Z_{A} =\left[2\left((s+b)\ln\left[\frac{(s+b)(b+\bsvar)}{b^2+(s+b)\bsvar}\right]-\frac{b^2}{\bsvar}\ln\left[1+\frac{\bsvar s}{b(b+\bsvar)}\right]\right)\right]^{1/2},
\end{equation}
where, for a given integrated luminosity, $s$ is the number of signal events, $b$ is the number of background events, and the uncertainty associated with the number of background  events is given by $\sigma_b$. 

\begin{figure}[!t]
\begin{center}
\centering
\includegraphics[scale=0.47]{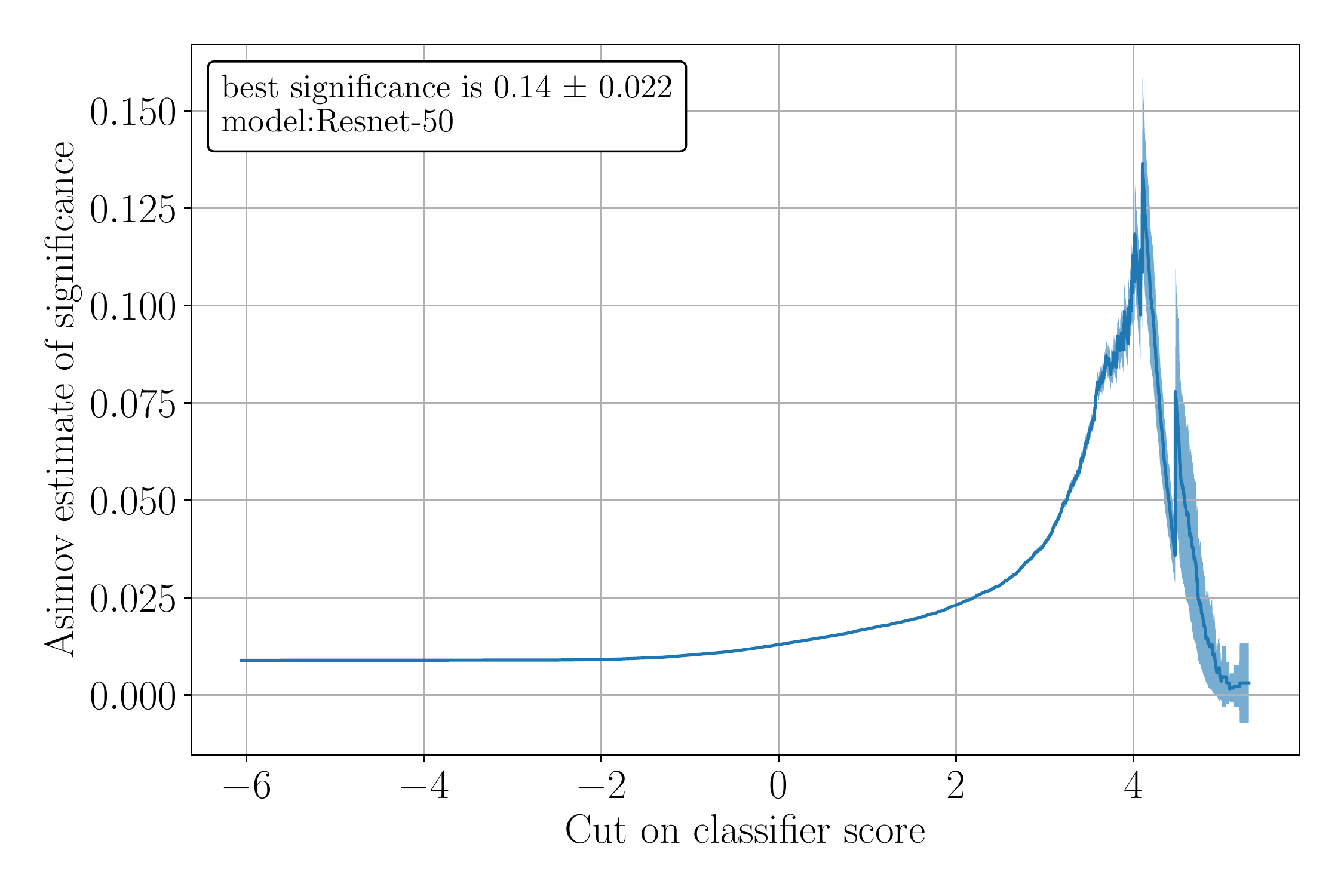} \\
\includegraphics[scale=0.85]{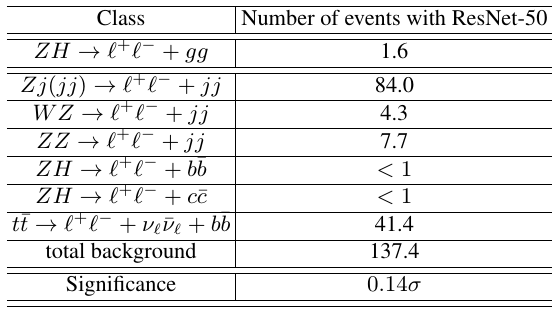} 
\caption{Upper panel: Asimov estimated significance for 3000 $\text{fb}^{-1}$, the best significance achieved by the ResNet-50 is around $0.14\sigma$. Lower panel: the number of signal and backgrounds events after imposing a CNN score cut of 4.2 in order to select the signal region. The last row displays the signal significance on the test set. The systematic error for these results is 0.33\% as in Ref.~\cite{Carpenter:2016mwd}. 
}
\label{ZA_1}
\end{center}
\end{figure}

In Fig.~\eqref{ZA_1}, upper panel, we estimate the Asimov significance for 3000 fb$^{-1}$ with a fixed 5\% systematic uncertainty, $\sigma_b/b$. The shaded blue band corresponds to a $\pm 1\sigma$ error band computed by propagating the statistical Poisson uncertainty of the background and signal counts~\cite{Elwood:2018qsr}. Our signal significance estimates therefore take statistical and systematic errors into account, even though in a simplified manner. As one can see, the ResNet-50, although gives an AUC of 0.91 for our signal, does not show promising prospects in terms of statistical significance due mainly to the large remaining number of events from the $Z$+jets and $t\bar{t}$ backgrounds. On the other hand, $ZZ$, $ZW$, $ZH(b\bar{b})$ and $ZH(c\bar{c})$ backgrounds are efficiently reduced by using abstract jet images. 

The number of signal and background events can be found in the table at the right panel of Fig.~\eqref{ZA_1}. Selecting events with scores larger than 4.2 gives the best signal significance at the expense of around two surviving signal events against more than 130 background events corresponding to a very small significance of $0.14\sigma$. Looking at the lower panel of Fig.~\eqref{ROC_1} we can easily understand the cut on the CNN score, it is a very hard cut to eliminate the backgrounds, relaxing it allows just too much backgrounds to increase the significance. The CNN actually does an excellent job in reducing the number of backgrounds but it is just not enough to give an useful signal significance to constrain the gluon jet decays.

\subsection{Classification with ResNet-50 and BDTs}

In spite of the fact that the CNN was not able to deliver a good signal significance, the classification scores of the signal and backgrounds can be used as a new feature to compose the data representation for some other ML classifier.
In order to further separate the signal samples from the backgrounds, we chose to stack the scores obtained from the CNN along with kinematic variables to construct another data representation to be classified by a boosted decision trees (BDT) algorithm. This type of ensemble has already been used to improve the classification power of particle physics analyses~\cite{Alves:2016htj,Aaltonen:2010jr}.




The representation of the data used to train the decision trees algorithm comprises the following variables:
\begin{itemize}
\item  two leptons invariant mass, $M_{\ell\ell}$; two jets invariant mass, $M_{jj}$; the invariant mass of the reconstructed leading jet, $M_{j}$; the invariant mass of the leading jet and the two leading leptons, $M_{j\ell\ell}$; the missing energy of the event $\not\!\! E_T,$ 

\item transverse momenta: $p^{\ell\ell}_{T},p^{j_{1}}_{T},p^{j_{2}}_{T},p^{b_{1}}_{T},p^{b_{2}}_{T},p^{\ell_{1}}_{T},p^{\ell_{2}}_{T}$,
\item angular distributions: the separation between the leading leptons pair and the leading jet, $\Delta R_{\ell\ell,j_1} = \sqrt{(\Delta \phi_{\ell\ell,j_1})^{2} + (\Delta \eta_{\ell\ell,j_1})^{2}}$; the separation in the azimuthal angle of the leading leptons, $\Delta\phi_{\ell,\ell}$; the cosine of the angle between the transverse momentum of the leading leptons pair and the leading lepton, $\cos(\Delta\phi_{\ell\ell,\ell_1})$; and between the leading leptons pair and the leading jet, $\cos(\Delta\phi_{\ell\ell,j_1})$, 

\item the $b$-tag of the event, this is 1 for events with at least one $b$-tagged jet, and 0 for events with no $b$-jets identified by \texttt{Delphes}

\item the score provided by the CNN model trained in the previous step described in Section~\ref{sec:CNN_model} for each one of the 7 classes. 
\end{itemize}

%
\begin{figure}[!t]
\begin{center}
\includegraphics[scale=0.35]{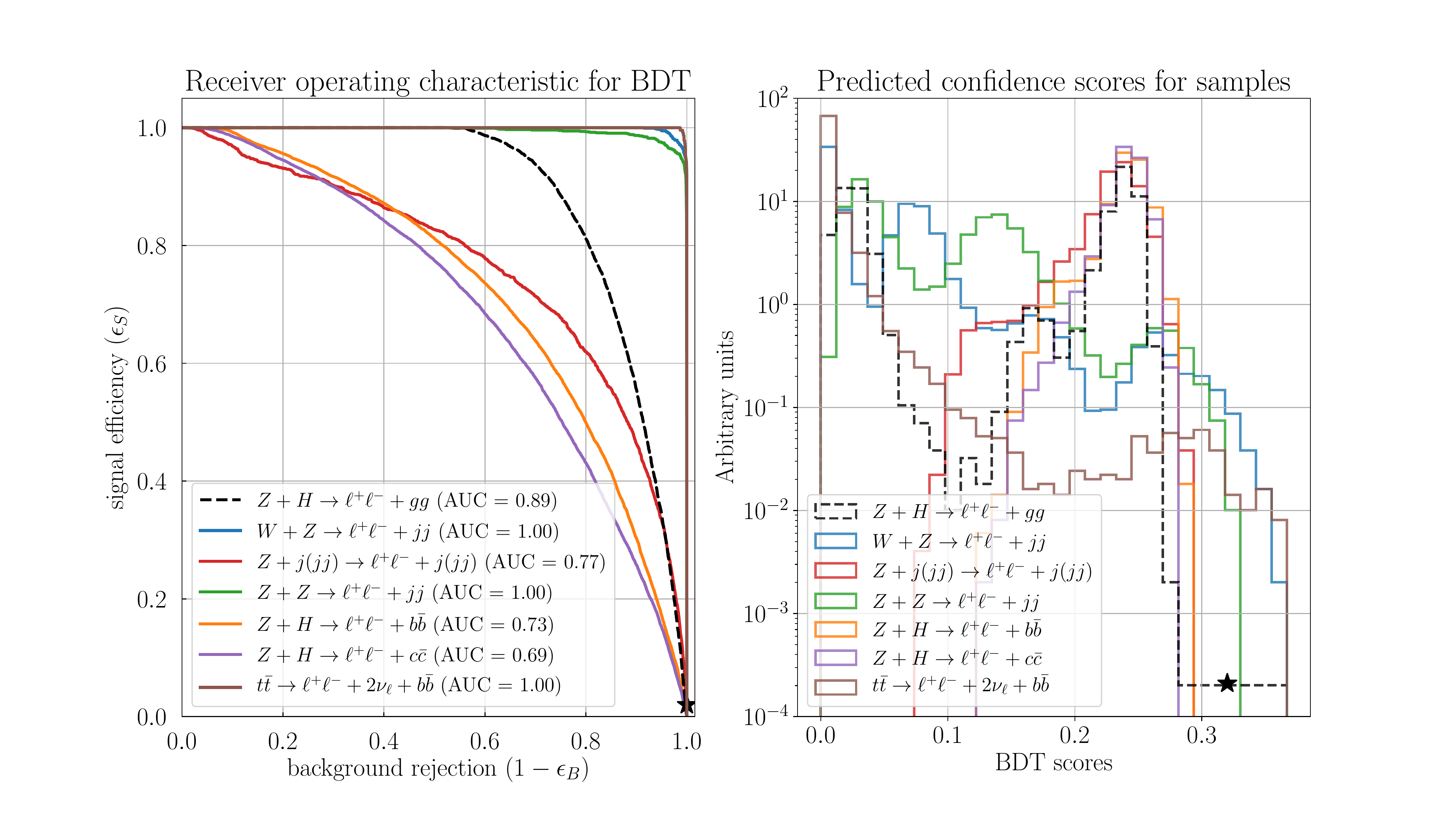} 
\caption{Left panel: ROC curves for the signal and backgrounds classes. Right panel: BDT scores for the signal and backgrounds classes. The star marker shows the cut selection that we impose to achieve the best Asimov significance given by this method.}
\label{ROC_2}
\end{center}
\end{figure}
For the BDT training and testing, we used the remaining 125k (40\%) events out of the original 312k events displayed in Table~\ref{tab:class_imagens}. Note that this is an independent data set not seen before by the CNN model. For each one of these events, we build a jet image and assign a CNN score with the ResNet-50 model trained in the previous step described in Section~\ref{sec:CNN_model}. We randomly split this data set as 80\% for training and 20\% for testing. We do not use synthetic images for the BDT train/test stage, just simulated ones. After constructing our new data set with the kinematic variables and CNN scores of each channel we can now turn our attention into the selection process of our ensemble classifier. A plethora of multivariate classification models are available right out of box in the \texttt{scikit-learn} \cite{scikit-learn} packages. Boosted decision tress are well-established classifiers and can provide very accurate predictions for classification tasks where the data set contains multiple classes.

BDTs, as any other ML algorithm, should be tuned in order to achieve a good classification performance.
A first approach is to use "brute force" to tune the hyperparameters by using a grid search but the number of combinations and the computational time to test each one of them increases exponentially. 
More efficient ways beyond grid search are random sampling or using gaussian process algorithms to learn the best hyperparameters. Another way to tackle this problem is to use genetic/evolutionary algorithms, as in Ref.~\cite{Freitas:2019hbk}. To do so, we make use of the Python Evolutionary Algorithm toolkit \texttt{DEAP}~\cite{DEAP_JMLR2012}, in conjunction with the the \texttt{scikit-learn} library. Such implementation cannot only provide the models with the highest accuracy, but also modify the "fitness" function of the evolutionary algorithm to search the best hyperparameters combination which returns the highest Asimov significance values. 

In our analysis, we found that a multi-class AdaBoost classifier~\cite{Zhu09multi-classadaboost} with 700 base estimators, a maximum tree depth of 5 and a learning rate of 1.0, keeping other hyperparameters as default options, presented the higher Asimov significance. Just like in the case of the CNN, we found that adjusting the hyperparameters to get a high accuracy does not guarantee the higher signal significance.

%
\begin{figure}[!t]
\begin{center}
\centering
\includegraphics[scale=0.47]{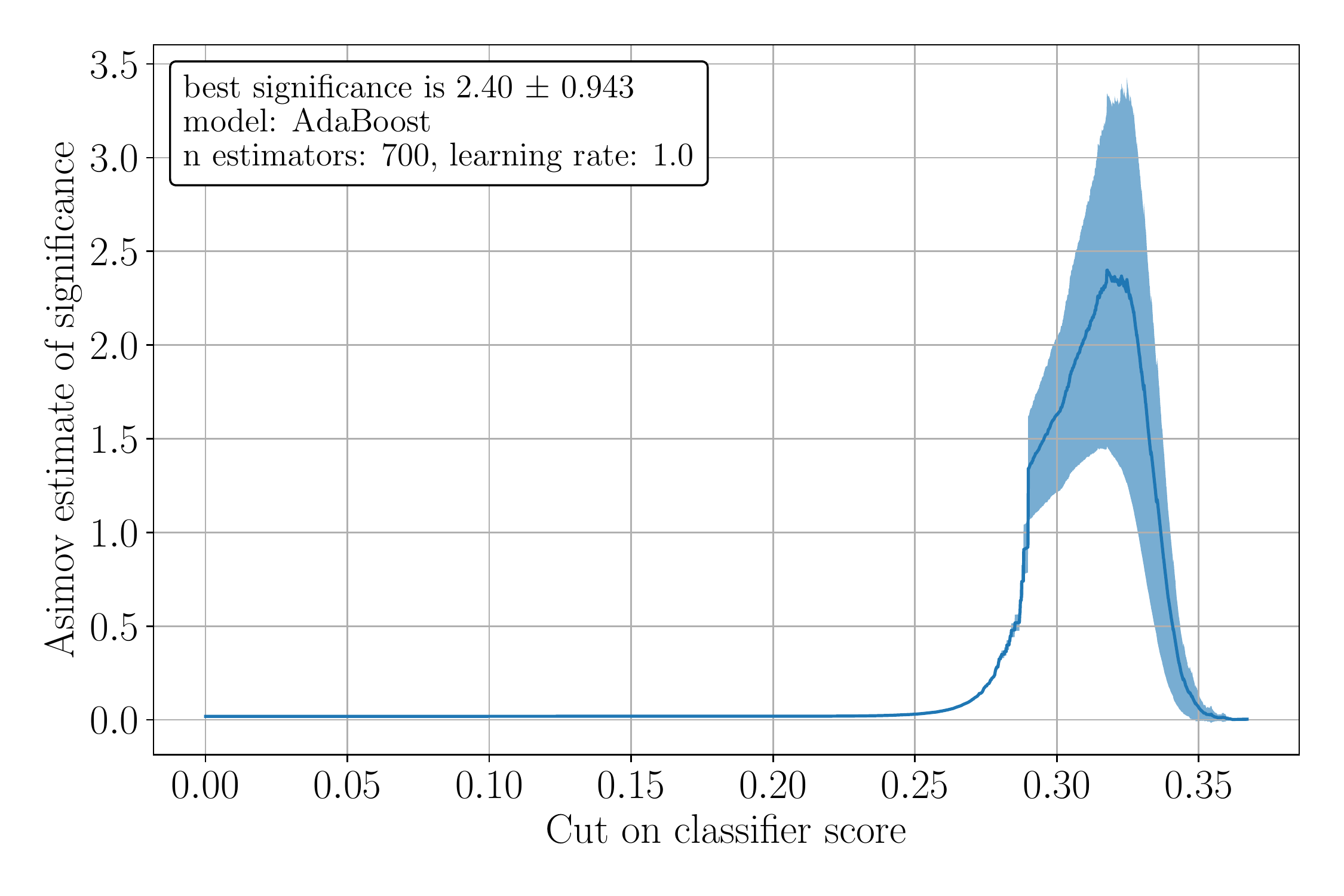} \\
\includegraphics[scale=0.8]{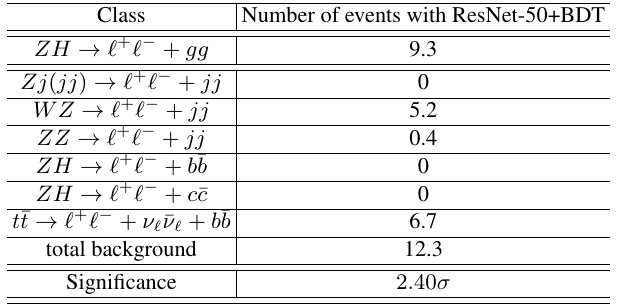} 
\caption{Upper panel: Asimov estimated significance for 3000 $\text{fb}^{-1}$, the best significance achieved by the BDT classifier is around $2.4\sigma$. Lower panel: the number of signal and background events after imposing a cut on the BDT score of 0.32. The last row displays the signal significance on the test set. The systematic error for these results is 0.33\% as in Ref.~\cite{Carpenter:2016mwd}.} 
\label{ASI_2}
\end{center}
\end{figure}
In Fig.~\eqref{ROC_2}, we show, at the left panel, the ROC curves after the the BDT classification. Remember that the CNN classification was very efficient to suppress all the backgrounds, except for $Z$+jets and $t\bar{t}$ backgrounds, but at the expense of cutting out too many signal events. Moreover, $Z$+jets and the contaminants $ZH(b\bar{b})$ and $t\bar{t}$ amounted to around 140 events resulting in a very small significance. The BDT found another solution instead, reducing $ZZ$ and $WZ$ by 50\% but completely eliminating the $Z$+jets and reducing $t\bar{t}$ by 85\% at the same time it keeps a much larger number of signal events. Comparing the ROC curves of Figs.~\eqref{ROC_1} and \eqref{ROC_2}, we clearly see that the AUC of the $Z$+jets increases very much. At the right panel of Fig.~\eqref{ROC_2}, we show the BDT scores of the event classes. The cut on the BDT score is also very hard but and the signal region is defined for a BDT score larger than 0.32. The star markers show the point of cut and the corresponding signal efficiency and background rejection.

In Fig.~\eqref{ASI_2}, we show, at the upper panel, the Asimov significance with its corresponding $\pm 1\sigma$ error band. The increase of the mean Asimov significance, $2.40\sigma$, assuming a 0.33\% systematic error, is quite pronounced compared to the CNN classification. Without any systematics, the signal significance is the same, $2.40\sigma$ and it barely changes even for a few percent systematic uncertainty due the $s/b$ ratio of 0.76. The sensitivity to systematics is considerably reduced compared to the cut-and-count analysis as shown in the next section. This is a consequence of including systematics at the tuning phase of the BDT algorithm -- the genetic algorithm is able, in this way, to find a configuration where the signal significance and the $s/b$ ratio are optimized together. This result greatly improves over the cut-and-count analysis of Ref.~\cite{Carpenter:2016mwd}, where a $0.25\sigma$ significance is achieved for the $ZH\to \ell^+\ell^- + gg$ channel with a signal-to-background ratio of order $10^{-4}$.

The much better result for BDT could raise doubts whether the CNN classification is necessary after all. To confirm that the CNN scores are playing any role in helping the BDT to separate the signals, we performed a feature importance analysis of the BDT features and found that indeed the CNN scores are the most important features for the BDT classification followed by $M_{\ell\ell}$, $\Delta\phi_{\ell\ell}$ and the missing transverse energy of the event. At the lower panel of Fig.~\eqref{ASI_2}, we show the number of events for each class after the cut on the BDT scores. The number of signal events jumps to around 15 events and the backgrounds by half compared to the CNN case. The $Z$+jets, $t\bar{t}$ and the $ZH$ contaminants are completely washed away, leaving just $ZZ$ and $WZ$ events in the backgrounds. 

We want to point out an important bonus in using ML algorithms to separate the signals from backgrounds in this process -- the $ZH(b\bar{b})$ and $ZH(c\bar{c})$ events are
suppressed with hard cuts on the score outputs of the classifiers. This makes the statistical analysis simpler and less prone to tagging efficiencies as we are going to show in the next section. 

\section{\label{sec:aaet} Signal significance and constraints on the light jet Higgs branching ratio}

With the results obtained in our ML analysis, we are able now to constrain the branching ratio of the Higgs boson into light jets. To do so, we closely follow the statistical analysis presented in Ref.~\cite{Carpenter:2016mwd}. It is important to highlight the differences between our analysis and the one performed in Ref.~\cite{Carpenter:2016mwd}. First of all, and more importantly, we consider the two-lepton category only, while in that work, the one+two-lepton categories are considered once they also take $WH$ into account in the analysis. Second, we include the signal contaminants $ZH(b\bar{b})$ and $ZH(c\bar{c})$ in the background category from the beginning. In Ref.~\cite{Carpenter:2016mwd}, these categories are considered only in the statistical analysis to constrain the light jets branching ratio multiplying the gluons signal
by suitable tagging and mistagging factors and, in this way, being able to discount for the $b\bar{b}$ and $c\bar{c}$ contamination. We, however, are able to increase the purity of
the signal by eliminating the $b$-jet and $c$-jet Higgs decays using our ensemble classifier as discussed in the previous section. After a hard cut on the BDT score, the number of bottom and charm jet contaminants is negligibly small, and only $ZZ$ and $ZW$ background events survive as discussed in the previous section. 

This efficient clearing up from $ZH(b\bar{b})$ and $ZH(c\bar{c})$ background events allows to place a direct upper bound on the Higgs to gluon jets branching ratio at the 95\% confidence level (CL)
\begin{eqnarray}
    && \mu_{j} = \frac{{\rm BR}(H\to gg)}{{\rm BR}^{\rm SM}(H\to gg)}\leq 1+\frac{\sqrt{\chi^2_{95\%}}}{S_j}=1.74\;(1.75)[1.78] \\ 
    && {\rm BR}(H\to gg) \leq 1.74(1.75)[1.78]\times {\rm BR}^{\rm SM}(H\to gg)\; ,
    \label{eq:chi2j}
\end{eqnarray}
where $S_j$ is the mean signal significance obtained  after the BDT classification computed with the simple significance metrics $S_j={s \over \sqrt{b+\sigma_b^2}}=2.7(2.6)[2.5]$, where $s=9.3$ and $b=12.3$, from the table at the right panel of Fig.~\eqref{ASI_2} assuming $\sigma_b/b=$ 0(5\%)[10\%] uncertainties in the background normalization. Note that these estimates might fluctuate in the blue band in Fig.~\eqref{ASI_2}. We use this simple significance metrics for a fair comparison with the results of Ref.~\cite{Carpenter:2016mwd} and because the results do not differ much compared with the Asimov formula.


%
\begin{table}[t!]
      \centering
      \caption{Flavor tagging efficiencies and the fraction of SM Higgs decay channels used to compute $\chi^2$ of Eq.~\eqref{eq:chi2}. Taken from Ref.~\cite{Carpenter:2016mwd}.}
      \label{tab:eff-channel}
\begin{tabular}{ cc }   
Flavor tagging efficiencies & Fraction of Higgs decay channels \\  
\begin{tabular}{ cccc } 
\hline
\hline
$\epsilon_{ai}$ & $b$-quark & $c$-quark & $j=g$ \\
$b$-tag & 70\% & 20\% & 1.25\% \\
$c$-tag & 13\% & 19\% & 0.50\% \\
un-tag $j'j'$ & 17\% & 61\% & 98.25\% \\
\hline
\hline
\end{tabular} &  
\begin{tabular}{ cccc } 
\hline
\hline
$e_{ai}$ & $b\bar b$ & $c\bar c$ & $jj$ \\
$bb$-tag & 99.6\% & 0.4\% & 0\% \\
$cc$-tag & 90.4\% & 9.6\% & 0\% \\
un-tag $j'$ & 16\% & 10\% & 74\% \\
\hline
\hline
\end{tabular} \\
\end{tabular}
\end{table}

We can also follow the steps of Ref.~\cite{Carpenter:2016mwd} to combine the signal strength obtained in our light jet analysis with other estimates taking into account tagging and mistagging factors that consider mixing among the jet classes. In this case, our analysis should be interpreted as a bound on untagged jets, that is it, jets that are not tagged
as $b$ or $c$-jets. In this case, we define the signal strength for a decay channel $H\to ii$ as
\bea
\mu_i = \frac{{\rm BR}(H\to i i)}{{\rm BR}^{\rm SM}(H\to i i)},
\eea
where we consider $ii=b\bar b,\ c\bar c,$ and $jj=gg$. Assuming each decay channel is statistically independent and following Gaussian statistics, we can get goodness of fit using the same chi-square function that we used in Eq.~\eqref{eq:chi2j} above
\begin{equation}
\begin{aligned}
\chi^2 &=  \sum_{a=j,c,b} \frac{(N_a-N^{\rm SM}_a)^2}{\sigma^2_a}\\
&= \sum_{a=j,c,b} {(\sum_{i=j,c,b} \epsilon_{ai}^2 {\rm BR}_i N_{\rm sig}^{\rm prod} - \sum_{i=j,c,b} \epsilon_{ai}^2 {\rm BR}_i^{\rm SM} N_{\rm sig}^{\rm prod})^2 \over (\sqrt{N_{\rm bkg}})^2} \\
&= \sum_{a=j,c,b} {(\sum_{i=j,c,b} e_{ai}\ \mu_i - 1)^2 \over (1/S_a)^2},\;\; e_{ai}=\frac{\epsilon^2_{ai} {\rm BR}_i}{\sum_{k=j,c,b} \epsilon^2_{ak} {\rm BR}_k}
\end{aligned}
\label{eq:chi2}
\end{equation}
where $S_a$ is the significance from each category identified by experiments, $\epsilon_{ai}$ are the tagging and mistagging factors, $e_{ai}=\epsilon$ are the fractions of the Higgs decay channel $i$ where both jets are tagged as $a$ given in Table~\ref{tab:eff-channel}, and $N_{\rm sig}^{\rm prod}$ is the number of signal events produced in the $ZH$ associated production.





\begin{figure}[t!]
  \center
  \includegraphics[width=0.45\linewidth]{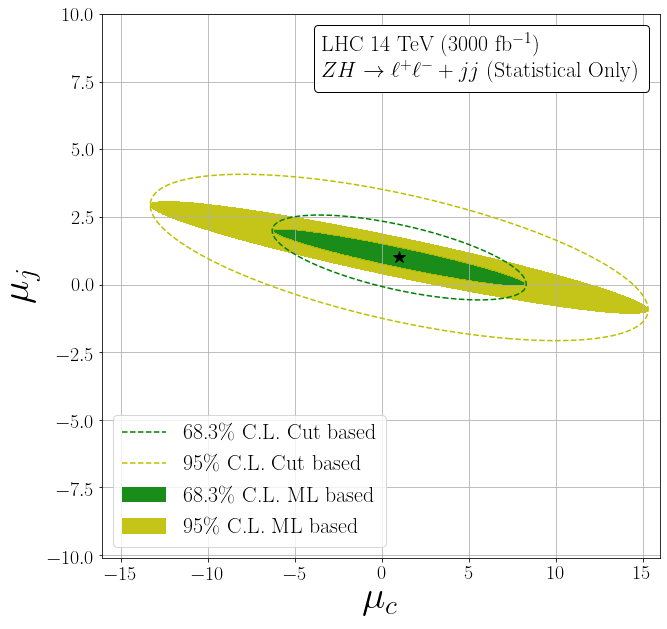}
  \includegraphics[width=0.45\linewidth]{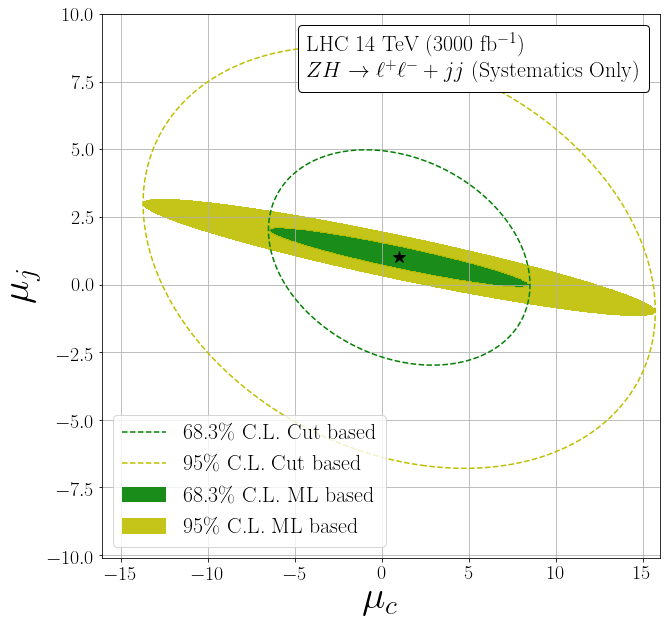}
  \caption{Contours in $\mu_c$-$\mu_j$ plane, for statistics only (left panel) and including systematic uncertainties (right panel) for the $\ell^+\ell^- + j'j'$ channel, respectively, where $j'$ is a jet that could not be tagged as $b$ or $c$-jet. The green and yellow regions are the 68.3\% and 95\% CL regions, respectively, given by the ML based analysis (combined BDT + ResNet-50 predictions), while the green and yellow dashed lines are the corresponding regions from the cut based analysis~\cite{Carpenter:2016mwd}.}
  \label{fig:ZH-contour}
\end{figure}

For the significance without systematic errors, we take $(S_b,S_c,S_j)$ = ($7.5,\ 1.35,\ 2.7$), while for the significance with 1\% of systematic errors we have $(S_b,S_c,S_j)$ = ($7.5,\ 1.35,\ 2.7$); in this case, the significances for $b$ and $c$ channels barely change. This level of systematic uncertainties seems to be adequate in view of the estimates of Ref.~\cite{Carpenter:2016mwd}. The $b$-channel significance $S_b$ = 7.5 comes from Table 12 in the ATLAS MC study \cite{ATL-PHYS-PUB-2014-011} for the category “Two-lepton”, while $c$-channel significance $S_c$ = $1.35$ comes from Fig.~2(a) of Ref.~\cite{Perez:2015lra} and computed in Ref.~\cite{Carpenter:2016mwd}. Contrary to $S_b$, $S_c$ includes contributions from $WH$ events then our estimate of the following signal strengths are approximated. 
The fully correlated signal strengths using only the  $\ell^+\ell^- + jj$ channel are plotted in Fig.~\eqref{fig:ZH-contour}.

We obtained the following 95\% CL upper bound limit for the Higgs branching ratio into untagged jets with 0\% and 1\% systematic errors, in parenthesis, for 3000 fb$^{-1}$
\begin{equation}
BR(H\rightarrow j'j') \leqslant 3.06 (3.10)\times BR^{SM}(H\rightarrow gg).
\end{equation}

The bound for ${\rm BR}(H\to c\bar c)$ can also be obtained from Eq.~\eqref{eq:chi2}. Actually, the contour plot in the $\mu_j\times \mu_c$ plane can be readily obtained as shown in Fig.~\eqref{fig:ZH-contour} which displays the cut-based contours from Ref.~\cite{Carpenter:2016mwd} as dashed lines and our results as shaded areas. We see that the constraints get considerably tighter for $\mu_j$ but not for $\mu_c$ which is expected once we are taking exactly the same significances for $b\bar b$ and $c\bar c$ signals as in the cut-based analysis. 

The signal-to-background ratio achievable after our classification is also significantly raised compared to cut-based analysis. This is important in order to get reliable significance estimates. The main sources of systematics are the total rates and the shapes of the kinematic distributions which can be affected by many features of the simulations, both for signals and backgrounds. The signal and background cross sections are impacted by missing higher order calculations. The inclusion of QCD K-factors is essential to better estimate these corrections. The factorization/renormalization scales impact both the total rates and also the shape of the distributions though. The simulation of hadronization and detector effects are also sources of systematic uncertainties. All these uncertainties should be taken into account in obtaining the output score of the ML algorithms that are used to perform the ultimate separation cut between the signal and background classes. Doing this analysis is however very demanding from the computational point of view once it is necessary to generate synthetic data for each variation of those parameters. 

Nevertheless, we can include systematics effects in the background yields in a simplified manner as we did in Eq.~\eqref{eq:asimov}. Although this is a limited account of systematic uncertainties, it is computationally feasible. It is also useful once our signal-to-background ratio is not too large. A good example about how several sources of systematic uncertainties can be taken into account in phenomenological studies using an ensemble of algorithms to boost the statistical significance of the signal can be found in Ref.~\cite{Aaltonen:2010jr}.

In the left panel of Fig.~\eqref{fig:ZH-contour} we show the case with no systematic uncertainties included, and the right panel shows how the bounds degrade once systematics in the background rates are taken into account. Even assuming a rather small systematics of 0.33\%, the cut-based bounds loosens compared to the statistics dominance case. On the other hand, the ML results do not change for this level of systematics once the $s/b$ ratio is much larger than that achieved with cut-and-count only. This is a preliminary indication that our results are more robust against systematic uncertainties.

\begin{table}[t!]
\centering
\begin{tabular}{c|c|c|c|c|c}
\hline
Systematics & 100 fb$^{-1}$ & 300 fb$^{-1}$ & 
1000 fb$^{-1}$ & 3000 fb$^{-1}$ & ATLAS+CMS combined(3 ab$^{-1})$\\ \hline\hline
0.33\% & $0.44\pm 0.16$ & $0.76\pm0.26$ & $1.38\pm0.48$ & $2.40\pm0.94$ & $3.38\pm 1.33$\\ \hline
1\% & $0.44\pm 0.16$ & $0.76\pm0.26$ & $1.38\pm0.48$ & $2.40\pm0.93$ & $3.38\pm1.33$\\ \hline
5\% & $0.44\pm 0.16$ & $0.76\pm0.26$ & $1.37\pm0.54$ & $2.35\pm1.00$ & $3.32\pm1.00$ \\ \hline
10\% & $0.44\pm 0.16$ & $0.75\pm0.30$ & $1.35\pm0.58$ & $2.23\pm1.08$ & $3.15\pm1.07$ \\ \hline
50\% & $0.41\pm 0.20$ & $0.65\pm0.32$ & $0.92\pm0.44$ & $1.10\pm0.60$ & $1.56\pm0.59$\\ \hline\hline
\end{tabular}
\caption[]{Projected Asimov significance of Eq.~\eqref{eq:asimov} for integrated luminosities of 100, 300, 1000 and 3000 fb$^{-1}$ at the 14 TeV LHC for the given systematic uncertainty after the BDT classification. In the last column we show the naive combination of both LHC experiments for an integrated luminosity of 3 ab$^{-1}$.
The uncertainties in the significance reflect the variation of the blue bands of Figure \eqref{ASI_2}.}
\label{tab:sys_vs_lumi}
\end{table}

In order to give an idea of the impact of systematic uncertainties, we compute the signal significance with our ML analysis with various levels of uncertainty and integrated luminosity. The results are shown in Table~\ref{tab:sys_vs_lumi}. First of all, we see that the results are rather robust against systematics levels up to 10\%. Second, 1000 fb$^{-1}$ should be enough to reach an $1\sigma$ significance, but only at the end of the HL-LHC run we should be able to probe Higgs to gluons in the $ZH\to \ell^+\ell^- + gg$ channel. Finally, in the last column we show the naive (added in quadrature) combination of data of both the LHC experiments after 3 ab$^{-1}$. Taking into account the statistical Poissonian uncertainty in the number of signal and background events and the systematics in the background yields, the LHC may reach a $3\sigma$ evidence for $pp\to ZH\to \ell^+\ell^- + gg$.

A caveat about these results is due. As we can observe in the right panel of Fig.~\eqref{ROC_2} and the upper panel of Fig.~\eqref{ASI_2}, a hard cut on the BDT output score is necessary to avoid backgrounds and achieve the maximum significance. We are thus probing the tail of the BDT score. Relaxing the cut to 0.3 adds statistics but at the price of a reduction of significance to an $\sim 1\sigma$ level. Moreover, systematic uncertainties on the shape of the features distributions might also change the tail of the output scores of the ML algorithms. As we discussed, evaluating the impact of these uncertainties is very computationally consuming.

In order to improve from the results reported in this work, $WH$ events can be used in combination to $ZH$, as in Ref.~\cite{Carpenter:2016mwd}. From the multivariate side of the analysis many other approaches can be employed: other kinds of jet taggers can be tested, different ensembles, meta-classifiers as new types of BDT and neural networks, and even a shape analysis of the output scores. In particular, a very promising direction can be attempted. This is the joint optimization of cuts, ML hyperparameters and jet finder parameters to construct jet images and boost the separation power of the meta-classifier, the BDT. This is, again, very time consuming. For each cut strategy, new jets should constructed and their images, then their output scores and the other kinematic features must feed the meta-classifier in order to have a final output score that is used to perform the final cut to separate the signals from the backgrounds. Previous applications of these joint optimizations are very encouraging~\cite{Alves:2019emf,Alves:2018oct,Alves:2017uls,Alves:2017ued}.

\section{\label{sec:conclusions} Conclusions}

Completing the SM picture in all its colors, shapes and nuances demands observing and studying all its predicted interactions.
The LHC is confirming that the Higgs boson coupling scales with the mass of the SM particles. This makes it easier to probe the Higgs couplings to bottom and top quarks, the heavy gauge bosons and the tau lepton but it turns the observation of interactions to the light quarks, especially up, down and strange quarks, and also the lighter leptons, much more challenging. Previous attempts to probe the Higgs coupling to light jets, that is it, to gluons and $u$, $d$ and $s$ quarks, based on standard cut-and-count strategies, have shown how difficult is to select the jets coming from the Higgs decay apart from the SM backgrounds.

In this work, we used machine learning tools to make the classification of signal and background jets more efficient. We found that analyzing $\ell^+\ell^- +jj$ events at the 14 TeV LHC, the decays from boosted $ZH$ production can be used to put stronger constraints on the branching ratio of $H\to gg$ than those of Ref.~\cite{Carpenter:2016mwd}. These boosted events generate fat jets whose substructure can be "seen" by convolutional neural networks after subtle transformations which make the jet images more discernible. We employed several state-of-art ML techniques to improve the performance of the CNN algorithm in obtaining the highest signal significance possible. In spite of its power, the CNNs were not able to separate signal from backgrounds at the level we need, however, the output scores assigned by the CNNs to each event class is by themselves a very distinctive feature that can be combined with kinematic information of the particles of the event to train another ML algorithm -- this is an {\it ensemble learning}. 

Following this insight, we trained a boosted decision trees algorithm using the CNNs outputs and kinematic information of the events to further cleaning of Higgs to gluon jets signals. We achieved a statistical significance for the signal class almost an order of magnitude larger compared to the cut-and-count analysis~\cite{Carpenter:2016mwd}, reaching $2.4\sigma$ in the statistics dominance scenario after 3000 fb$^{-1}$. Assuming a rather optimistic systematic uncertainty on the background rate of just 0.33\%, the cut analysis presents even more difficult prospects of $0.1\sigma$ for the channel $ZH\to\ell^+\ell^- + gg$ alone, according to Ref.~\cite{Carpenter:2016mwd}. Our results, on the other hand, are much more robust against systematic uncertainties in the background rates, with $2.4\sigma$, an order of magnitude improvement over the cut analysis. By the way, even a 10\% systematics on the backgrounds normalization should not change these results considerably.  Naively combining the data from both CMS and ATLAS and taking into account the variation in the ML performance in our cross validation, the LHC may find evidence at
the $3\sigma$ level for light jet decays of the Higgs boson.

Moreover, the ML algorithm was able to eliminate the $Z(H\to b\bar{b})$ and $Z(H\to c\bar{c})$ contaminants allowing us to derive the following 95\% CL bound directly on the light jets branching ratio instead of a bound on the untagged jet class as in Ref.~\cite{Carpenter:2016mwd} 
$$
{\rm BR}(H\to gg)\leq 1.74(1.78)\times {\rm BR}^{SM}(H\to gg)\; ,
$$
assuming a 0(10)\% systematic uncertainty on the background normalization.

Combining the significance reached in this analysis with the ones in the search for $H\to b\bar{b}$ and $H\to c\bar{c}$ taking into account mixing of tagged and mistagged jet classes, it is possible to put bounds on the ${\rm BR}(H\to j'j')$, where $j'$ is a jet that could be not associated to a $b$ ot $c$-jet, it is an "untagged" jet. Following closely the analysis of Ref.~\cite{Carpenter:2016mwd}, we found
$$
{\rm BR}(H\to j'j')\leq 3.06 (3.10)\times {\rm BR}^{SM}(H\to gg)\; ,
$$
for 0(1)\% of systematics, which improves the results obtained exclusively with a dedicated cut-and-count analysis.

The results presented in this work can be improved in several ways. For example, events from $WH$ can be included in the analysis. Many other ML algorithms can also be tested, including other types and levels of ensembles. Tuning ML hyperparameters and kinematic cuts jointly also has the potential to greatly improve the signal significance as demonstrated in Refs.~\cite{Alves:2019emf,Alves:2018oct,Alves:2017uls,Alves:2017ued}. More kinematic features can also be incorporated in the data representation seeking to strengthen the correlations of the classes. We believe, however, that any significant improvement will come from the machine learning side of the analysis. Discovering the SM Higgs to light jets can only be possible for a future collider, unless new physics intervenes to enhance these couplings. In this case, a dedicate study can benefit from our results and analyses.

\section*{Acknowledgments}
AA thanks Conselho Nacional de Desenvolvimento Cient\'{\i}fico (CNPq) for its financial support, grant 307265/2017-0. FFF is partially supported by the China Postdoctoral Science Foundation project Y8Y2411B11 and the project \textit{From Higgs Phenomenology to the Unification of Fundamental Interactions} PTDC/FIS-PAR/31000/2017 grant BPD-32 (19661/2019). FFF thanks Prof. C. Herdeiro and Prof. A. P. Morais for the hospitality during his stay at Aveiro university. AA would like to thank Tilman Plehn, Gregor Kasieczka, Juan Gonzales-Fraile and Anja Butter and for helpful discussions in the early stages of this project.

\appendix

\section{Glossary of terms}\label{glossary}

We present here a minimal glossary of machine learning and data science terms to help the reader to capture the key ideas of the work concerning the construction of the algorithms. Several books and texts can be further explored to a proper understanding of the many details contained in this work. We, particularly, recommend the following ones of Refs.~\cite{MEHTA20191,Abdughani:2019wuv,annurev.nucl.012809.104427,albertsson2018machine,RevModPhys.91.045002,Nature560} to an introduction to ML for physicists, and the following ones~\cite{HAN_2006,Langacker:1995hi,Pich:2012sx} for ML experts that wish to acquire the basics of particle physics phenomenology.

\begin{itemize}
\item[]   {\bf True positive rate ({\it tpr }$\equiv$ $\epsilon_S$) : }  ratio of true positive count and total signal events. The true positive counts are
 the number of signal events correctly identified by the algorithm. It also corresponds to the usual notion of signal acceptance.
\item[] {\bf False positive rate ({\it fpr }$\equiv$ $\epsilon_B$) :}  ratio of false positive counts and total number of background events.  The false positive counts are the background events, predicted as signal events by the algorithm. It also corresponds to the usual notion of  background rejection.
\item[] {\bf Receiver operating characteristic (ROC) : }  plot of {\it tpr} as a function of {\it fpr}   for each value of the classifier threshold between 0 to 1. 
\item[] {\bf Area under the curve (AUC) : }area under the ROC curve and a typical measure of the algorithm performance. 
\item[] {\bf Accuracy :} ratio of the correctly identified signal and background events versus total number of signal and background events.
\item[] {\bf Learning curve :} curve with shows the performance of the algorithm with iterative runs i.e. behaviour of the loss function with iterations. 
\item[] {\bf Batch :} data is divided into small sets, called batches, to save time and computation efforts.
\item[] {\bf Hidden Layers :} intermediate layers between  the input and output layers.
\item[] {\bf Loss function :} the function which the algorithm searches to minimise. 
\item[] {\bf Epochs :} The period between initialisation of the search for the minimum and when the batches pass  the NN. Basically, number of epochs is an 
iteration counter of how many times complete data set is explored by the algorithm, such that learning parameters are optimized.
\item[] {\bf Dropout: } mechanism to avoid the model overfitting, whereby the NN could drop few of the units (neurons) at the time of training.
\item[] {\bf Pretraining :} Quick pre-run with smaller number of epochs and steeper loss functions. The longer training is initialised by the pretraining hyperparameters.  
\item[] {\bf Classifier output :}  set of predictions for test sample. Our analysis is a binary classification problem, so with the previously decided (user-decided) classification threshold, the events will either belong to signal or background class.
\end{itemize}

\section{Label Smoothing Cross Entropy}\label{labelsmoothie}
The last layer of our ResNet model is a fully-connected layer with a size being equal to the number of classes, denoted by $c$, to output the predicted confidence scores. Given an abstract image from our data set, denote by $z_i$, the predicted score for class $i$. These scores can be normalized by the softmax operator to obtain predicted probabilities. Denote by $q$ the output of the softmax operator $q=\text{softmax}(z)$, the probability for class $i$, $q_i$, can be computed by:

\begin{equation}
q_i = \frac{\exp(z_i)}{\sum_{j=1}^c \exp(z_j)}\; ,
\end{equation}
where $q_i > 0$ and $\sum_{i=1}^c q_i = 1$, so $q$ is a valid probability distribution. 

On the other hand, assume the true label of this image is $y$, we can construct a truth probability distribution to be $p_i = 1$ if $i=y$ and 0 otherwise. During training, we minimize the negative cross entropy loss

\begin{equation}
\ell(p, q) = - \sum_{i=1}^c q_i \log p_i
\end{equation}
to update model parameters in order to make these two probability distributions similar to each other. In particular, by the way how $p$ is
constructed, we know that $\ell(p, q) = - \log p_y = - {z_y} + \log\left(\sum_{i=1}^c \exp(z_i)\right)$. The optimal solution is $z_y^*\to \infty$ while keeping others small enough. In other words, it encourages the output scores dramatically distinctive which potentially leads to overfitting.

The idea of label smoothing was first proposed to train Inception-v2~\cite{DBLP:journals/corr/SzegedyVISW15}. It changes the
construction of the true probability to
\begin{equation}
q_i =
\begin{cases}
  1-\varepsilon & \quad\textrm{if } i = y, \\
   \varepsilon / (c-1) & \quad\textrm{otherwise,}\\
\end{cases}
\end{equation}
\noindent where $\varepsilon$ is a small constant. Now the optimal solution becomes

\begin{equation}
z_i^* =
\begin{cases}
  \log( (c-1)(1-\varepsilon)/\varepsilon) + \alpha & \quad\textrm{if } i = y, \\
   \alpha & \quad\textrm{otherwise,}\\
\end{cases}
\end{equation}
\noindent where $\alpha$ can be an arbitrary real number. This encourages a finite output from the fully-connected layer and can generalize better.

\begin{figure}[!t]
\begin{center}
\includegraphics[width=0.32\textwidth]{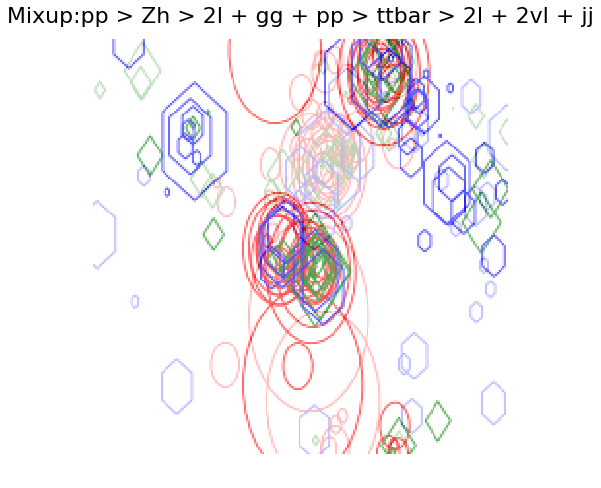} 
\caption{Mixup result for a given signal ($ZH\to \ell^{+}\ell^{-} + gg$) and background ($t\bar{t}$) abstract images. The images are combined according to the value $\lambda=0.8$, that is it, in the image shown we have a new image that corresponds to 80\% of the signal image and 20\% of the background image.}
\label{fig_mixup}
\end{center}
\end{figure}

\section{MixUp}\label{mixup}

The \textit{mixup} training \cite{DBLP:journals/corr/abs-1710-09412} consists  of, during the training phase, randomly sample two images $(x_i, y_i)$ and $(x_j, y_j)$ from our data set. Then we form a new image by a weighted linear interpolation of these two:

\begin{eqnarray}
\hat x &=& \lambda x_i + (1-\lambda) x_j, \\
\hat y &=& \lambda y_i + (1-\lambda) y_j,
\end{eqnarray}

\noindent where $\lambda \in [0, 1]$. Therefore, \textit{mixup} extends the training distribution by incorporating the prior knowledge that linear interpolations of feature vectors should lead to linear interpolations of the associated targets. We show an example in Fig.~\eqref{fig_mixup}.

\bibliography{main_arxiv}
\end{document}